\documentclass[aps,twocolumn,prl,amsmath,showpacs,superscriptaddress]{revtex4}
\usepackage{epsfig}
\usepackage{bm}
\usepackage{graphicx}
\def\be{\begin{equation}}
\def\ee{\end{equation}}
\def\bea{\begin{eqnarray}}
\def\eea{\end{eqnarray}}

\def\r#1{(\ref{#1})}
\def\nn{\nonumber\\}

\begin{document}

\title{Effective forces induced by fluctuating interface: exact results.}
\author{D. B. Abraham}
\affiliation{Theoretical Physics, Department of Physics, University of
  Oxford, 1 Keble Road, Oxford OX1 3NP, United Kingdom} 
\affiliation{Department of Chemistry and Miller Institute for Basic
  Science, University of California, Berkeley, USA}
\author{F. H. L. Essler}
\affiliation{Theoretical Physics, Department of Physics, University of
  Oxford, 1 Keble Road, Oxford OX1 3NP, United Kingdom} 
\author{A. Macio\l ek }
\affiliation{Theoretical Physics, Department of Physics, University of
  Oxford, 1 Keble Road, Oxford OX1 3NP, United Kingdom} 
\affiliation{Max-Planck-Institut f{\"u}r Metallforschung,
  Heisenbergstr.~3, D-70569 Stuttgart, Germany} 
\affiliation{Institute of Physical Chemistry, Polish Academy of Sciences,
Department III, Kasprzaka 44/52, PL-01-224 Warsaw, Poland}

\date{\today}
\begin{abstract}

We present  exact derivations of the effective capillary wave
fluctuation induced forces resulting from pinning  of an interface
 between two coexisting phases at two points  separated by a distance  $r$.
 In two dimensions   the  Ising ferromagnet calculations
based on  the transfer matrix approach give an attractive force decaying
as $1/r$ for large distances. In three dimensions mapping of the body-centered
solid-on-solid model onto the 6-vertex model allows for exact solution
using the  bosonization analysis of the equivalent 
 XXZ Heisenberg quantum chain.
The exact result gives the  attractive force which  decays asymptotically as 
$1/(r\log r)$. 

\end{abstract}
\maketitle

Interfaces between coexisting thermodynamic phases exhibit large spatial
fluctuations at temperatures above the roughening transition. 
The length scale of such fluctuations diverges with system size, so that
the interface is ``washed out'' in the thermodynamic limit
 (unless suitable external fields are applied), a result completely 
at variance with classical thermodynamical expectations. Nevertheless, 
such a striking phenomenon has recently been observed by a direct visual method
in an ingeniously constructed experiment with a colloidal system
\cite{science04}. Further, there are many implicit manifestations in
such diverse areas as biological systems \cite{bio}, the asymptotic
behaviour of correlation functions \cite{cor} and vicinal sections of
crystals in the terrace-ledge-kink (TLK) model \cite{TKL,FiFi}.
Here ledge fluctuations are examined, typically
when there is a short range interaction between such ledges, either attractive
or repulsive, together with a long-ranged repulsion which comes about because
neighbouring ledges cannot cross and so restrict each others' configuration
space. Such a force is said to be of ``entropic'' or ``Casimir'' type.
A crucial fact which will be used later in another context, 
is that the TLK model of a crystal surface can be mapped onto the 6-vertex
model of Lieb \cite{Lieb} and solved exactly, with various phase transitions
resulting, depending on the nature of the short-ranged force: {\it attraction}
gives surface reconstruction, {\it repulsion}
facet selection at a ``magic'' angle, with
the facet disappearing by a roughening mechanism.
In this Letter we consider a related interfacial problem, but one in which 
geometric limitation of the configuration of the interface manifests itself
in a long range {\it attraction}. 
This we analyse in $d=2$ and $d=3$ by using exact
$d=2$ Ising techniques in the former case and another 6-vertex
mapping, together with some new results, in the latter case. This
6-vertex mapping, due to van Beijeren \cite{BCSOS}, is different from
the TLK one, but has the crucial common feature that it gives height
conservation around closed loops,  a necessary condition for the
integrity of the surface. Further, it manifests a phase transition of
roughening type \cite{rough}. 

It is well-known that chemico-physical forces can trap colloidal particles
at the liquid-liquid or liquid-vapour interface at coexistence
\cite{pieran:80:0}. This phenomenon  is
widely used, for example, to create the stable emulsions in systems
called Pickering emulsions \cite{pickering}.
Another scenario here  is that of
a bifunctional particle, say a sphere, the surface of which is
divided at the equator into two hemispheres with differential wetting
properties. 
Since the colloidal particles are much larger and more massive than the 
ones in the phase separating systems, they restrict the motion of the interface
to which they are rigidly attracted strongly. As a limiting case, they
can be treated as spatially fixed. The phase space  
restriction then implies, as we shall show later, that, were the massive 
particles to be released, then they would accelerate towards each other.
The interest of such questions stems not only from 
the various  practical  applications 
\cite{pieran:80:0,joann:01:0,dinsm:02:0}  but also from the basic
 requirement to
understand the  nature of the effective forces  between the
colloidal particles \cite{chen:05:0,sear:99:0}. In particular, the
important role of the aforementioned capillary fluctuations is clear;
 substantial progress has recently been made
\cite{oettel:05:0,lehle:06:0}, but there remain a number of issues.

\begin{figure}[htb]
\includegraphics[scale=0.38]{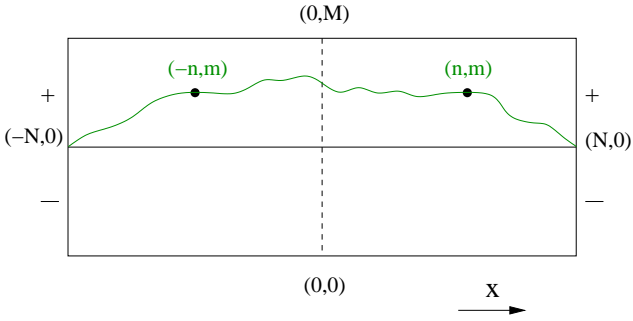}
\caption{Schematic plot of Ising  lattice with  interface pinned 
by its endpoints and in two  interior point
at the same level.
The nearest-neighbor ferromagnetic coupling is $K$in unit of $kT$.
 The transfer matrix acts in $x$ direction. }  
\label{fig0}
\end{figure}
\vspace*{-0.4cm}
First, we calculate the incremental free energy of an interface in the
$d=2$ Ising ferromagnet resulting from pinning, which in this case can
be achieved in various ways. The method, which carries over to our d=3
calculations, is to fix the interface at its extremities and to use
this for the reference state. The pinned state is then a further
restriction of the reference state in which two interior points are
required to have the same interface displacement, but all such
displacements are allowed with equal weight. The associated partition
functions are calculated using the standard transfer matrix technique,
supplemented with the ``domain wall state method'' \cite{7}: the state
$|m\rangle$ achieves localization at the position $m$ on any line
normal to the mean interface direction and can be written as
\begin{equation}
\label{eq:domst}
|m\rangle=M^{-1/2}\sum_k \exp -ik(m+1/2)\ \vartheta(k)\ G^\dagger(k)|\Phi\rangle.
\end{equation}
Here $|\Phi\rangle$ is both the $G(k)$ vaccum and the maximal eigenvector 
of the transfer matrix $V$. The $G^\dagger(k)$ are Fermi
operators which ``diagonalize'' the transfer matrix for a lattice of
width $M$ normal to the transfer direction \cite{LSM}. 
Thinking of the transfer matrix as generating a discrete Euclidean time
evolution, the $G^\dagger(k)$ create ``particles'' with energy
$\gamma(k)$ given by Onsager's formula
$\cosh\gamma(k)=\cosh 2K\cosh 2K^*-\cos k$,
where $K$ is the nearest neighbour coupling in units of $k_BT$,
$K^*$ is the dual coupling given by the involution
$\sinh(2K)\sinh(2K^*)=1$.  It is important to note that
$|m\rangle$ only localises the interface up to the bulk correlation
length as is revealed by calculating the local magnetization \cite{7}. The
ratio of the partition function with two pinned  states $|m\rangle$ 
separated by a horizontal distance $2n$ (see Fig.~\ref{fig0}), denoted by $Z^{\times}(n|N,M)$
 to  that for the unpinned system, denoted by $Z(N,M)$, is
\bea
\label{eq:3}
\frac{Z^{\times}(n|N,M)}{Z(N,M)}&=&\frac{1}{\langle 0|V^{2N}|0\rangle}
\sum_m\langle 0|V^{N-m}|m\rangle\nn
&\times&
\langle m|V^{2n}|m\rangle\langle m|V^{N-m}|0\rangle
\eea
As $M\to\infty$, bringing in expression for $\langle m|$ gives
\be
\label{eq:sum}
\frac{Z^{\times}(n|N,\infty)}{Z(N,\infty)}=
\sum_m \frac{\bigl[U(N-n,m)\bigr]^2 U(2n,m)}{U(2N,0)}\ ,
\ee
where
\be
U(N,m)=\int_0^{2\pi}\frac{d\omega}{2\pi}\
e^{-N\gamma(\omega)+im\omega}\ .
\ee
Evaluating the sum over $m$ in (\ref{eq:sum}) gives
\be
\frac{Z^{\times}(n|N,\infty)}{Z}=
\frac{U(2N-2n,0)\ U(2n,0)}{U(2N,0)}\ .
\ee
Finally, using the fact that
\be
\lim_{N\to \infty}\frac{U(2N-2n,0)}{U(2N,0)}= e^{-2n\gamma(0)}\ ,
\ee
we conclude that in the limit $M, N\to \infty$ the incremental
 free energy is given by
\be
\beta\ f^\times(n)=-\log\left[
\int_{0}^{2\pi}\frac{d\omega}{2\pi} e^{-2n[\gamma(\omega)-\gamma(0)]}
\right].
\label{eq:7}
\end{equation}
 Here $\gamma(0)$ is proportional to
the inverse correlation length. The asymptotics for large $n$ of this are determined by the Laplace method
\be
\beta\ f^\times\sim \log\sqrt{2\pi \gamma^{(2)}(0)r/a_0}+{\cal O}(1),
\label{eq:7c}
\ee
where $r=2na_0$ and $a_0$ is the lattice spacing, valid  for $n\gamma^{(2)}(0)\gg 1$, i.e., $r\gg a_0/\gamma ^{(2)}(0)$, where $1/\gamma ^{(2)}(0)$ is
 the  interface stiffness. The stiffness comes in because the surface tension
is angle dependent on the lattice. The question of angle dependence of 
(\ref{eq:7c}) will be discussed elsewhere.
The implied force is
\be
{\cal F}(r)\sim -\frac{kT}{2}\frac{1}{r}.
\label{eq:7b}
\ee
Notice that the law itself does not depend on the stiffness, 
or on surface tension, but the range of validity does. Techniques like
those in Ref.~\cite{science04} might be used to make the lower bound on $r$
 as small as possible.
The reader might feel that the pinning mechanism is somewhat
contrived. If the interface were localized both at the boundaries and
at both the interior points, perhaps a more intuitive restriction
than ``floating'' at the same level, then $Z^\times/Z\to 0$ as
$N\to\infty$, rendering the definition useless. In experiments one 
always has a finite system, which may introduce an additive term in
\r{eq:7}, which is independent of $n$, but which may well diverge with
$N$, as would be the case with the alternative pinning just
described. But for any finite $N$, this would not come up in the force
as in \r{eq:7b}. There are many variants here, which will be
developed in a longer paper.

\begin{figure}[htb]
\includegraphics[scale=0.38]{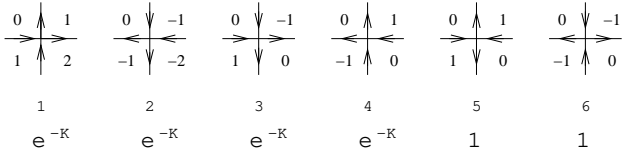}
\caption{The arrow configurations of allowed configurations of the 6-vertex
model. The rule for associating plaquette heights on the dual lattice is that,
pointing along the arrow, the plaquette on the right is higher. 
Vertices 1 to 4 have a diagonal height change in one or the other
direction. Vertices 5 and 6 have a ripple (no diagonal height change)
and are deemed to be flat. The weight assignment, with $K>0$, favors
rippling over diagonal height change. }  
\label{fig}
\end{figure}

We turn now to $d=3$  and consider the van Beijeren BCSOS \cite{BCSOS}
model, where height changes are associated with vertex configurations 
as shown in Fig.~\ref{fig}. The fact that the allowed configurations 
are those of the $6-$vertex model  with two 
arrows in and two arrows out, guarantees height conservation around
any close loop, which is required for the integrity of the sheet.
Along the $(1,0)$ direction, which is normal to the transfer direction, the height
difference of abutting plaquettes is strictly $\pm 1$. Labelling
the row of vertical arrows by eigenvalues of $2S_j^z$ for spin~-1/2,
 the following XXZ Heisenberg - Ising Hamiltonian 
\be
{\cal H}_{XXZ}=J\sum_{j=1}^M \left[(S^x_jS^x_{j+1}+S^y_jS^y_{j+1})+\Delta S^z_jS^z_{j+1}\right]\ ,
\label{HXXZ}
\ee
$S_{M+1}^{\alpha}=S_1^{\alpha}, \alpha = x, y, z$ (cyclic boundary conditions)
commutes with the transfer operator, provided  $\Delta=1-(1/2)e^{K}$. Thus
$-\infty< \Delta < 1/2$.
The region of interest here is for $-1<\Delta <1/2$ where
 the interface is {\it rough}.
The ground state of (\ref{HXXZ}) furnishes the maximum eigenvalue
of the transfer matrix. Proceeding in the $(1,0)$ direction, 
the height difference
at a separation $r=2na_0$ for integer $n$ is 
\be
\delta h=2\sum_{j=1}^{2n} S_j^z\ .
\label{heightd}
\ee
In this letter, we are interested in the situation when $\delta h=0$.
The ratio of partition function with this height fixing  to that without it
is
\be
\frac{Z^{\times}(n)}{Z}=\int_0^{\pi}\frac{d\theta}{\pi}f(\theta,2n)=e^{-\beta f^{\times}(n)}\ ,
\ee
where $f^{\times}(n)$ is the associated incremental free energy and
\be
f(\theta,2n)=\langle \exp\left(i\theta\sum_{j=1}^{2n}
S^z_j\right)\rangle_{{\cal H}_{XXZ}}.
\ee
We are interested in the large-$n$ asymptotics of $f(\theta,2n)$.
We note that $\exp(-\beta f^{\times}(n))$ is simply the probability 
that the $z$-component of
the spin on the interval $[1,2n]$ is zero. In this sense this quantity  is
related to Korepin's ``emptiness formation probability'' and its
generalizations \cite{korepin,abanov}.
To calculate $\exp(-\beta f^{\times}(n))$ we employ a bosonization analysis.

It is well known that the large distance behaviour of correlation
functions in the Heisenberg-Ising model \r{HXXZ} is described by a
Gaussian model \cite{gauss}
\be
{\mathcal H}=\frac{1}{16\pi}\int dx \left[v
(\partial_x\Phi)^2+\frac{1}{v}(\partial_t\Phi)^2\right].
\label{GaussianModel}
\ee
Here the scalar field $\Phi$ is compact
$\Phi(x)\equiv \Phi(x)+8\pi\alpha$,
and the velocity $v$ and compactification radius $4\alpha$ are given
in terms of the parameters of the lattice Hamiltonian \r{HXXZ} and the
lattice spacing $a_0$ as
$\alpha=\left({\rm arccos}(-\Delta)/4\pi\right)^{(1/2)}$, and
$v=(Ja_0\sin(4\pi\alpha^2))/(2(1-4\alpha^2))$.
The lattice spin operator $S^z_j$ is expressed in terms of the bosonic
field as
\bea
S^z_j&\sim&\frac{a_0}{8\pi\alpha}\,\partial_x\Phi(x)\nn
&-& A(-1)^ja_0^{1/8\alpha^2}
\sin\left(\frac{\Phi(x)}{4\alpha}\right)+\ldots\ ,
\label{bosoSz}
\eea
where $x=ja_0$ and $A$ is a known constant \cite{Lukyanov}. We now
turn to the bosonization of $\exp\Bigl(i\theta\sum_{j=1}^{n}
S^z_j+\frac{1}{2}\Bigr)$. We first note that this operator is 
a $2\pi$-periodic function of $\theta$, which allows us to write
\bea
f(\theta,n)&=&\frac{e^{-in\theta}}{2}\Biggl[\Bigl
\langle e^{i\theta\sum_{j=1}^{n}\bigl(S^z_j+\frac{1}{2}\bigr)}
\Bigr\rangle_{{\cal H}_{XXZ}}\nn
&+&\Bigl\langle
e^{i(\theta-2\pi)\sum_{j=1}^{n}\bigl(
S^z_j+\frac{1}{2}\bigr)}\Bigr\rangle_{{\cal H}_{XXZ}}\Biggr].
\eea
Application of \r{bosoSz} then gives (the staggered part of $S^z_j$
does not contribute to the sum as $\Phi(x)$ is a slowly varying field)
\bea
e^{i\theta\sum_{j=1}^{n}S^z_j}
&\sim&A(\theta)
\exp\Biggl(\frac{i\theta}{8\pi\alpha}\int_0^{na_0}\!\!\!dx\
\partial_x\Phi\Biggr) \nn
&=&A(\theta)
\exp\biggl(\frac{i\theta}{8\pi\alpha}[\Phi(na_0)-\Phi(0)]\biggr).
\label{vertexop}
\eea
Here $A(\theta)$ is a normalization constant. The expectation value of
\r{vertexop} is easily calculated
\be
\Bigl\langle e^{i\theta\sum_{j=1}^{n}S^z_j}\Bigr\rangle_{{\cal H}}\sim
A(\theta)\ (nc)^{({\theta}/{4\pi\alpha})^2},
\ee
where $c^{-1}a_0$ is a short-distance cutoff. Using this result we
obtain the following expression for $f(\theta,n)$
\bea
f(\theta,n)&\sim&
\frac{A(\theta)}{2}(mc)^{-(\theta/4\pi\alpha)^2}\nn
&+&(-1)^{n} \frac{A(\theta-2\pi)}{2}(nc)^{-((\theta-2\pi)/4\pi\alpha)^2}.
\label{result}
\eea
We can check \r{result} by comparing it to the known expression
at the free fermion point $\Delta=0$. Here $f(\theta,n)$ can be
expressed as a Toeplitz determinant, see \cite{LSM,XY} and references
therein.  The large-$n$ asymptotics have been calculated in
\cite{ovchinnikov} with the result 
\bea
f(\theta,n)=\frac{a(\theta)}{(2n)^\frac{\theta^2}{2\pi^2}}
+(-1)^{n}\frac{a(1-\theta)}{(2n)^\frac{(\theta-2\pi)^2}{2\pi^2}}\ ,
\eea
where
\be
a(\theta)=\exp\bigl(\frac{\theta^2}{2\pi^2}\int_0^\infty\frac{dt}{t}
\bigl[e^{-2t}-\frac{4\pi^2}{\theta^2}\frac{\sinh^2(\theta t/2\pi)}{\sinh^2(t)}\bigr]
\bigr).
\ee
This agrees with the general expression \r{result}.
It is now a simple matter to calculate $\exp\bigl(f^{\times}(2n)\bigr)$, as for
large $n$ the $\theta$-integral is dominated by its saddle point. Hence 
\be
e^{-\beta f^{\times}(n)}\sim\frac{{\cal A}}{\sqrt{\log(nc')}}\ ,
\label{ef2}
\ee
where ${\cal A}$ and $c'$ are ($\Delta$-dependent) constants.
Thus, the resulting force is attractive and behaves for large $r$ as
\be
{\cal F}(r)\sim -\frac{kT}{2r\log(rc'/a_0)}\ .
\label{force2}
\ee
Notice the difference between \r{ef2} and \r{eq:7b}.

In this Letter, we have presented exact calculations of the incremental free
energy resulting from pinning the interface between two coexisting phases
at two points separated by a distance $r$. The case for $d=2$ is more detailed
because it allows the interface to have a diffusive structure at the molecular
level, as is known to be the case in real systems. The resulting force is attractive with a remarkably long-ranged decay as $1/r$. In $d=3$, we have treated 
this pinning phenomenon within the BCSOS model of a random surface, using
van Beijeren's  isomorphism with the 6-vertex model. Thus, the interface 
is locally sharp. It is defined at points on a $Z^2$ square lattice. 
The attractive force in this case does not differ strongly from 
the $d=2$ one, in that it decays as $1/(r\log r)$ instead of as $1/r$, 
a rather striking result.
Note that these forces are of much longer range than the electrostatic ones
induced by photon fluctuations, termed van der Waals \cite{vdw} or the 
fluctuation-induced Casimir-type repulsion between extended objects
 in similar systems. This $1/(r\log r)$ result has also been obtained by
Lehle, Oettel and Dietrich \cite{lehle:06:0} in a continuum model for
a particular, but very physically reasonable case of contact of the interface
with two {\it extended} objects; these we may think of as spheres each with
two hemispherical surfaces having different wetting properties which keep
the sphere at the interface at an orientation and immersion which makes the
equatorial line separating the two wetting region sit exactly in the interface,
as mentioned at the beginning of this Letter.
The case we have considered is an extreme one in that the colloidal sphere
has been treated as though it were a point (or a unit cube on the dual
lattice). In mitigation, our results for d=3 are obtained for
mesoscopic modelling which retains, unlike the usual capillary wave
scenarios, molecular scale discreteness and which also shows a
roughening transition.

\acknowledgments
This work was  supported in part by the EPSRC under grant
EP/D050952/1. DBA thanks the Miller institute for financial support
and  D.C. Chandler and K.B. Whaley for interesting discussions
and hospitality at UC Berkeley. AM thanks  S. Dietrich for valuable
comments.


\begin{thebibliography}{99}
\bibitem{science04} 
D.G.A.L. Aarts, M. Schmidt, and H.N.W. Lekkerkerker, Science {\bf 304}, 847 (2004).
\bibitem{cor} D. B. Abraham, Phys. Rev. Lett. {\bf 50}, 291 (1983).
\bibitem{TKL} D. B. Abraham, Phys. Rev. Lett. {\bf 51}, 1279 (1983);
 D. B. Abraham, F. H. L. Essler and F. T. Latrem\'oli\`ere,
 Nucl. Phys. B {\bf 556}, 411 (1999).  
\bibitem{FiFi}
M.E. Fisher and D.S. Fisher, Phys. Rev. B{\bf 25}, 3192 (1982).
\bibitem{bio} 
E. Sackmann and R.F. Bruinsma, Chem. Phys. Chem. {\bf 3}, 262 (2002).
\bibitem{BCSOS} H. van Beijeren, Phys. Rev. Lett. {\bf 38}, 993 (1977). 
\bibitem{rough}
J.D. Weeks, G.H. Gilmer and H.J. Leamy, Phys. Rev. Lett. {\bf 31}, 549 (1973).
\bibitem{pieran:80:0} P. Piera\' nski, Phys. Rev. Lett. {\bf 45}, 569 (1980).
\bibitem{pickering} S. U. Pickering, J. Chem. Soc. {\bf 91}, 2001 (1907). 
\bibitem{Lieb} E. H. Lieb, F. Y. Wu, 
{\it Two-dimensional Ferroelectric Models}, in Phase Transitions
 and Critical Phenomena, Vol.1, p. 331, ed. by C. Domb and J. Lebowitz
 (Academic London, 1972). 
\bibitem{joann:01:0} J. D. Jannopoulos, Nature (London), {\bf 414}, 257 (2001).
\bibitem{dinsm:02:0} A. D. Dinsmore, M. F. Hsu, M. G. Nikolaides, M. M\'arquez, A. R. Bausch, and D. A. Weitz, Science {\bf 298}, 1006 (2002).
\bibitem{chen:05:0} W. Chen, S. Tan, T.-K. Ng, W. T. Ford and P. Tong, Phys. Rev. Lett. {\bf 95}, 218301 (2005).
\bibitem{sear:99:0} R. P. Sear, S.-W. Chung, G. Markovich, W. M. Gelbart, and J. R. Heath, Phys. Rev. E {\bf 59}, R6255 (1999).
\bibitem{oettel:05:0} M. Oettel, A. Dominguez, and S. Dietrich, J. Phys.: Condens. Matter, {\bf 17}, L337 (2005).
\bibitem{lehle:06:0} H. Lehle, M. Oettel, and S. Dietrich, Europhys. Lett.,
 {\bf 75}, 174 (2006).
\bibitem{7}  D. B. Abraham and F. T. Latrem\'oli\`ere, Phys. Rev. Lett.
 {\bf 77}, 171 (1996).
\bibitem{onsager} L. Onsager, Phys. Rev. {\bf 65}, 117 (1944).
\bibitem{LSM}
T.D. Schultz, D.C. Mattis and E.H. Lieb, Rev. Mod. Phys. {\bf 36}, 856 (1964).
\bibitem{golest:00:0} R. Golestanian, Phys. Rev. E, {\bf 62}, 5242 (2000).
\bibitem{korepin}
V. E. Korepin, A. G. Izergin and N. M. Bogoliubov, {\em {Quantum Inverse
  Scattering Method, Correlation Functions and Algebraic Bethe Ansatz}}
  (Cambridge University Press, 1993).
\bibitem{abanov}
A.G. Abanov and V.E. Korepin, Nucl.Phys. {\bf B647}, 565 (2002).
\bibitem{gauss}
A. Luther and I. Peschel, Phys. Rev. B {\bf 12}, 3908 (1975);
F.D.M. Haldane, Phys. Rev. Lett. {\bf 47}, 1840 (1981);
I. Affleck, {\rm in} {\em Fields, Strings and Critical Phenomena},
{\rm eds E. Br\'ezin and J. Zinn-Justin},(Elsevier, Amsterdam, 1989).
\bibitem{Lukyanov}
S. Lukyanov, Phys. Rev. B{\bf 59} 11163 (1999).
\bibitem{XY}
E. Lieb, T. Schultz and D. Mattis, Ann. Phys. {\bf 16}, 407 (1961);
B.M. McCoy, Phys. Rev. {\bf 173}, 531 (1968);
E. Barouch and B.M. McCoy, Phys. Rev. A {\bf 3}, 786 (1971);
F. Colomo, A.G. Izergin, V.E. Korepin and V. Tognetti,
Theor. Mat. Phys. {\bf 94}, 19 (1993);
M. Shiroishi, M. Takahashi and Y. Nishiyama,
J. Phys. Soc. Jap. {\bf 70}, 3535 (2001);
N. Kitanine, J.M. Maillet, N.A. Slavnov and V. Terras,
Nucl.Phys. {\bf B642}, 433 (2002);
\bibitem{ovchinnikov}
A.A. Ovchinnikov, math-ph/0509026
\bibitem{vdw} J. N. Israelachvili, {\it Intermolecular and Surface Forces}, 2nd ed. Academic Press (1992).
\end{thebibliography}
\end{document}